\documentclass[a4paper,11pt]{appolb}
\usepackage{amsmath}
\usepackage{graphicx}
\usepackage{appendix}
\usepackage{xcolor}

\begin{document}

\title{Progress in implementing the kinematical constraint into the small-x JIMWLK evolution equation
}

\author{P. Korcyl\thanks{Presented by P.K. at the XXX Cracow Epiphany Conference on Precision Physics at the High Energy Colliders, Jan 8 - 12, 2024}, L. Motyka and T. Stebel
\address{Institute of Theoretical Physics, Jagiellonian University, ul. \L ojasiewicza 11, 30-348 Krak\'ow, Poland}
}
\maketitle

\begin{abstract}
    The most complete high-energy evolution of Wilson line operators is described by the set of equations called Balitsky-JIMWLK evolution equations. It is known from the studies of the linear - the BFKL - evolution equation that the leading corrections come from the kinematically enhanced double collinear logarithms. A method for resumming such logarithmic corrections to all orders for the Balitsky-Kovchegov equation is known under the name of kinematical constraint. In this work, we discuss the progress in implementing these corrections into the Langevin formulation of the JIMWLK equation. In particular, we introduce a set of correlation functions which are nonlocal in the rapidity variable. They appear in the construction of the kinematical constraint, however, their behavior with rapidity has not been investigated numerically so far. We derive their large-$N$ evolution equations, solve them numerically, and comment on their implications for the implementation of the full kinematical constraint. 
\end{abstract}

\section{Introduction}
\label{sec:intro}

In Deep Inelastic Scattering experiments on protons, such as the ones conducted at HERA accelerator or planned at the EIC collider in Brookhaven National Laboratory, the QCD dynamics in the high-energy limit is believed to be captured by an effective description called Color Glass Condensate \cite{CGC1_iancu2002colour,CGC2_Iancu_2004,CGC3_JALILIANMARIAN_2006,CGC4_WEIGERT_2005,CGC5_gelis2006lectures,CGC6_GELIS_2007,CGC7_Gelis:2010nm}. It is founded on the observation that in the saturation regime, the dense and slow gluons of the target can be described by classical fields and are separated in energy from the fast and energetic gluons from the projectile. 
This framework allows to calculate various measurable observables, for instance, the di-jet cross-section in the nearly back-to-back limit \cite{Caucal:2023nci,Caucal:2023fsf}. The crucial element of such calculation is the evolution equation which relates measurements at different values of the Bjorken $x$ variable. The most general evolution equation is the so-called Balitsky-JIMWLK set of equations \cite{JIMWLK1_Balitsky_1996,JIMWLK2_Jalilian_Marian_1997,JIMWLK3_Jalilian_Marian_1998,JIMWLK4_Kovner_2000,JIMWLK5_Iancu_2001,JIMWLK6_Iancu_2001,JIMWLK7_Ferreiro_2002}. In the limit of a large number of color, $N_c$, a simplified, closed equation for the dipole gluon amplitude can be derived, called the Balitsky-Kovchegov equation \cite{JIMWLK1_Balitsky_1996,BK_Kovchegov_1999}. For some observables, like the mentioned di-jet cross-section in the nearly back-to-back limit, the knowledge of the dipole amplitude is not enough and one should also use more complicated correlation functions like the Weiz\"acker-Williams distribution. The latter can only be obtained by solving the full B-JIMWLK equation. The purpose of this contribution is to describe our efforts to include subleading corrections to the numerical framework for solving the leading order B-JIMWLK equation. 

\subsection{Leading order}

The leading order B-JIMWLK equation is expressed in terms of Wilson lines $U(\mathbf{x})$ at a position $\mathbf{x}$ of the transverse plane of the collision. It is numerically convenient to employ its formulation as a stochastic Langevin equation \cite{Langevin1_Weigert_2002,Langevin2_Blaizot_2003,Langevin3_Rummukainen_2004}. The target rapidity is discretized in small steps $\epsilon$, $\eta_n = n \epsilon$, and the rapidity $n\epsilon$ of a Wilson line is denoted by the lower subscript, $U_{n\epsilon}(\mathbf{x})$. Thus, the Wilson line at the next step of the evolution, $\eta_{n+1}=(n+1)\epsilon$ is given by \cite{Lappi:2012vw}
\begin{equation}
\label{eq. JIMWLK}
U_{(n+1)\epsilon}(\mathbf{x}) = \exp\Big( i \sqrt{\epsilon} \alpha^L_{n+1}(\mathbf{x})\Big) U_{n\epsilon}(\mathbf{x}) \exp\Big(-i \sqrt{\epsilon} \alpha^R_{n+1}(\mathbf{x}) \Big)
\end{equation}
where
\begin{align}
\label{eq. alpha}
    \alpha_{n+1}^L(\mathbf{x}) &= \frac{1}{\pi} \int_{\mathbf{z}} \sqrt{\alpha_s} K^i_{xz} \xi^i_{n+1}(\mathbf{z}),\\
    \alpha_{n+1}^R(\mathbf{x}) &= \frac{1}{\pi} \int_{\mathbf{z}} \sqrt{\alpha_s} U^{\dagger}_{n\epsilon}(\mathbf{z}) K^i_{xz} \xi^i_{n+1}(\mathbf{z}) U_{n\epsilon}(\mathbf{z}),
\end{align}
and the Langevin noise is hidden in the two-dimensional vector of matrix-valued variables $\xi^i(\mathbf{z})$.  These random vectors are generated anew on each site of the lattice in each step of the evolution,
\begin{equation}
\boldsymbol{\xi}(\mathbf{x}) = (\xi^x(\mathbf{x}), \xi^y(\mathbf{x})) = (\xi^x_a(\mathbf{x}) t^a, \xi^y_b(\mathbf{x}) t^b) \,,
\end{equation}
from a normal distribution with unit width. The vectors $\boldsymbol{\xi}(\mathbf{x})$ are uncorrelated in $\mathbf{x}$, therefore
\begin{equation}
\label{eq. noise}
\langle \xi_a^i(\mathbf{x})\, \xi_b^j(\mathbf{y}) \rangle = \delta_{ab} \delta^{ij}\, \delta(\mathbf{x} - \mathbf{y}) \,. 
\end{equation}
The leading order kernel $K^i_{xz}$ is given by
\begin{equation}
    K^i_{xz} = \frac{(\mathbf{x} - \mathbf{z})^i}{(\mathbf{x} - \mathbf{z})^2}.
\end{equation}
The numerical studies of this formulation of the JIMWLK equation have been performed by several groups \cite{Langevin3_Rummukainen_2004,SOL1_Lappi_2011,SOL2_Dumitru_2011,SOL3_Lappi_2013,SOL4_Schlichting_2014,SOL5_Schenke_2016,Korcyl:2020orf,Cali:2021tsh,Korcyl:2021pef} and it was shown that the dynamics of the dipole amplitude can be reproduced by the large-$N_c$ BK equation. Typically, in the numerical framework the transverse plane is discretized into a lattice of points spaced by $a$ in each direction and of the finite total volume, i.e. the extent in each direction is set to $L=Na$, with $N$ natural. In order to keep translational symmetry periodic boundary conditions are imposed, effectively turning the plane into a torus.

\subsection{Kinematical constraint}

The kinematically enhanced subleading corrections were first discussed in Refs.\cite{KC1_Andersson:1995ju, KC2_Kwiecinski:1996td,KC3_Salam:1998tj}. Their implementation in the BK equation was proposed in Ref.\cite{KC4_Golec-Biernat:2001dqn} and subsequently, their position space representation was presented in Refs.\cite{KC5_Motyka:2009gi,KC6_Beuf:2014uia,KC7_Iancu:2015vea}. Later, it was proposed in Ref.\cite{Hatta:2016ujq} to implement the kinematical constraint in the framework described above by introducing a couple of modifications to the Wilson lines $U(\mathbf{x})$ and to the $\alpha_{n}^L(\mathbf{x})$ and $\alpha_{n}^R(\mathbf{x})$ variables while maintaining the overall structure of the Langevin equation Eq.\eqref{eq. JIMWLK}. The Wilson lines receive an additional index $r$ indicating the scale of the very first dipole,
\begin{equation}
    U(\mathbf{x}, \eta) \rightarrow U(\mathbf{x}, r, \eta),
\end{equation}
where $r$ is the distance at which the dipole correlation function $S(r, \eta)$,
\begin{equation}
S(r, \eta) = \frac{1}{N_c} \langle \langle \textrm{tr} U^{\dagger}(\mathbf{x}, r, \eta) U(\mathbf{x+r}, r, \eta) \rangle_{\mathbf{x}} \rangle_{\textrm{CGC}}
\label{eq. dipole amplitude}
\end{equation}
is being evaluated. Here, we use translational symmetry to average over all possible sites $\mathbf{x}$ and $\langle \dots \rangle_{\textrm{CGC}}$ denotes the typical Color Glass Condensate averaging over the initial color sources configurations. Subsequently, 
\begin{align}
    \label{eq_new_alphaL}
    \alpha_{n+1}^L(\mathbf{x}) \rightarrow \alpha_{n+1}^L(\mathbf{x},r) &= \frac{1}{\pi} \int_{\mathbf{z}} \sqrt{\alpha_s} \theta(n \epsilon - \delta^r_{r_{xz}} ) K^i_{xz} \xi^i_{n+1}(\mathbf{z}),\\
    \label{eq_new_alphaR}
    \alpha_{n+1}^R(\mathbf{x}) \rightarrow \alpha_{n+1}^R(\mathbf{x},r) &= \frac{1}{\pi} \int_{\mathbf{z}} \sqrt{\alpha_s} \theta(n \epsilon - \delta^r_{r_{xz}} ) U^{\dagger}(\mathbf{z}, R^r_{xz}, n\epsilon - \delta^r_{r_{xz}}) \times \nonumber \\ &\times K^i_{xz} \xi^i_{n+1}(\mathbf{z}) U(\mathbf{z}, R^r_{xz}, n\epsilon - \delta^r_{r_{xz}}).
\end{align}
where
\begin{equation}
    \delta^r_{r_{xz}} = \max \{0, \ln \frac{r^2}{(\mathbf{x} - \mathbf{z})^2} \},
\end{equation}
\begin{equation}
    R^r_{xz} = \max \{r, |\mathbf{x} - \mathbf{z}| \}.
\end{equation}
Note that these equations differ from similar expressions presented in Ref.\cite{Hatta:2016ujq} where the evolution was described in terms of the dipole rapidity $Y$. Here we follow Refs.\cite{ETA1_Ducloue:2019ezk,ETA2_Ducloue:2019jmy} and work in the target rapidity variable $\eta$.

The effect of the modifications in Eqs.(\ref{eq_new_alphaL}, \ref{eq_new_alphaR}) is that the $\theta$ function cuts out a part of the phase space in the convolution. At $\eta=0$ it removes a disc of radius $r$ around the site $\mathbf{x}$ which then shrinks for $\eta >0$. Additionally, Eq.\eqref{eq_new_alphaR} contains Wilson lines at a shifted rapidity, i.e. when $\delta^r_{r_{xz}}>0$ it involves Wilson lines from the \emph{earlier (past)} stages of the simulation.

The numerical implementation of Eqs.(\ref{eq_new_alphaL}, \ref{eq_new_alphaR}) into the JIMWLK equation Eq.\eqref{eq. JIMWLK} yields unexpected results since in some cases the obtained evolution of the dipole amplitude with the kinematical constraint is \emph{faster} than the evolution at leading order. The purpose of the rest of this contribution is to try to identify the possible reasons for such behavior.

\subsection{Recovering BK equation}
\label{sec. recovering}

Taking the definition of $S(r,\eta+\epsilon)$ from Eq.\eqref{eq. dipole amplitude}, expanding for small $\epsilon$ and using Eq.\eqref{eq. JIMWLK} for each Wilson line individually is expected to yield an evolution equation for the dipole amplitude $S(r,\eta)$ with the kinematical constraint \cite{Hatta:2016ujq}. This turns out not to be a straightforward step and one has to make additional assumptions. Here, we concentrate on a new class of correlation functions which appear at these intermediate stages and which contain Wilson lines at different values of rapidity. It is our suspicion that the unexpected behavior of these functions may corrupt the convergence of the results to the known solutions of the BK equation with kinematical constraint.

\section{Correlation functions nonlocal in the target rapidity}

In order to simplify the presentation in the following we discuss only two representatives of the correlation functions nonlocal in the target rapidity. We denote them by $C$ and $W$. Moreover, at the first step, we study their behavior with the leading order evolution equation, hence we use the Wilson lines $U(\mathbf{x})$ as building blocks, and their evolution is given by Eq.\eqref{eq. JIMWLK}. Although such correlation functions appear in the construction of the kinematical constraint in Ref.\cite{Hatta:2016ujq}, to our knowledge it is the first time that they have been evaluated numerically and studied in the framework of leading order JIMWLK equation.

\subsection{Definitions}

The rapidity correlator $C(\mathbf{x}, \eta)$ is defined as
\begin{equation}
\label{eq. rapidity correlator}
    C(\mathbf{x}, \eta) = \frac{1}{N_c} \langle \textrm{tr} U^{\dagger}(\mathbf{x}, 0) U(\mathbf{x}, \eta) \rangle_{\textrm{CGC}}.
\end{equation}
In many cases we equivalently consider the volume averaged version, $C(\eta) = \frac{1}{V} \sum_{\mathbf{x}} C(\mathbf{x},\eta)$ which, due to the maintained translational symmetry, should be equal.
Similarly, we define the correlator $W(r,\eta)$
\begin{equation}
\label{eq. W correlator}
    W(\mathbf{r},\eta) = \frac{1}{N_c} \frac{1}{V} \sum_{\mathbf{x}} \langle \textrm{tr} U^{\dagger}(\mathbf{x}, 0) U(\mathbf{x}+\mathbf{r}, \eta) \rangle_{\textrm{CGC}}.
\end{equation}
Obviously, $W(0,\eta) \equiv C(\eta)$ as well as $W(r,0) \equiv S(r,0)$. 

\subsection{Large-$N_c$ limit}

In order to establish the dependence on $\eta$, we expand $C(\eta+ \epsilon)$ and $W(r,\eta+ \epsilon)$ in $\epsilon$,
\begin{align}
    C(\mathbf{x}, \eta+\epsilon) &= \frac{1}{N_c} \langle \textrm{tr} U^{\dagger}(\mathbf{x}, 0) U(\mathbf{x}, \eta+\epsilon) \rangle, \\
    W(\mathbf{r},\eta+\epsilon) &= \frac{1}{N_c} \langle \textrm{tr} U^{\dagger}(\mathbf{x}, 0) U(\mathbf{x}+r, \eta+\epsilon) \rangle.
\end{align}
Since only one of the Wilson lines depends on $\eta$ the derivation is much simpler than that for $S(r,\eta)$. After some algebra we obtain a set of equations
\begin{align}
    \frac{\partial W(\mathbf{x}-\mathbf{y},\eta)}{\partial \eta} 
    &= \frac{\bar{\alpha}_s}{2 \pi} \int_{\mathbf{z}} \mathcal{K}_{xz}\Big(  S(\mathbf{x}-\mathbf{z},\eta) W(\mathbf{z}-\mathbf{y},\eta) - W(\mathbf{x}-\mathbf{y},\eta) \Big) \label{eq. W}\\
    \frac{\partial C(\eta)}{\partial \eta} 
      &= \frac{\bar{\alpha}_s}{2 \pi} \int_{\mathbf{z}} \mathcal{K}_{xz} \Big(
        S(\mathbf{x}-\mathbf{z},\eta) W(\mathbf{z}-\mathbf{x},\eta)  -C(\mathbf{x}, \eta) \Big) \label{eq. C}\\
    \frac{\partial S(\mathbf{x}-\mathbf{y},\eta))}{\partial \eta} &=  \frac{\bar{\alpha}_s}{2 \pi}    \int_{\mathbf{z}} \mathcal{M}_{xyz} \Big(
      S(\mathbf{x}-\mathbf{z},\eta) S(\mathbf{z}-\mathbf{y},\eta) - S(\mathbf{x}-\mathbf{y},\eta)\Big) \label{eq. normal bk eq}
\end{align}
Several comments are in order now. First, Eq.\eqref{eq. normal bk eq} is the well-known leading order BK equation for the dipole amplitude. It is a closed equation that involves only $S$ so it can be solved without any reference to the other correlation functions. That is not the case for the other two equations, Eqs.(\ref{eq. W}, \ref{eq. C}), which mix $C$, $W$, and $S$. In principle, $C$ and $W$ may be complex, however, we see that if the initial condition is real, then the evolution does not introduce imaginary parts, therefore the correlation functions $W$, $C$, and $S$ always remain real in this approximation.

\subsection{Initial slope}

\begin{center}
\begin{figure}
    \includegraphics[width=1.0\textwidth]{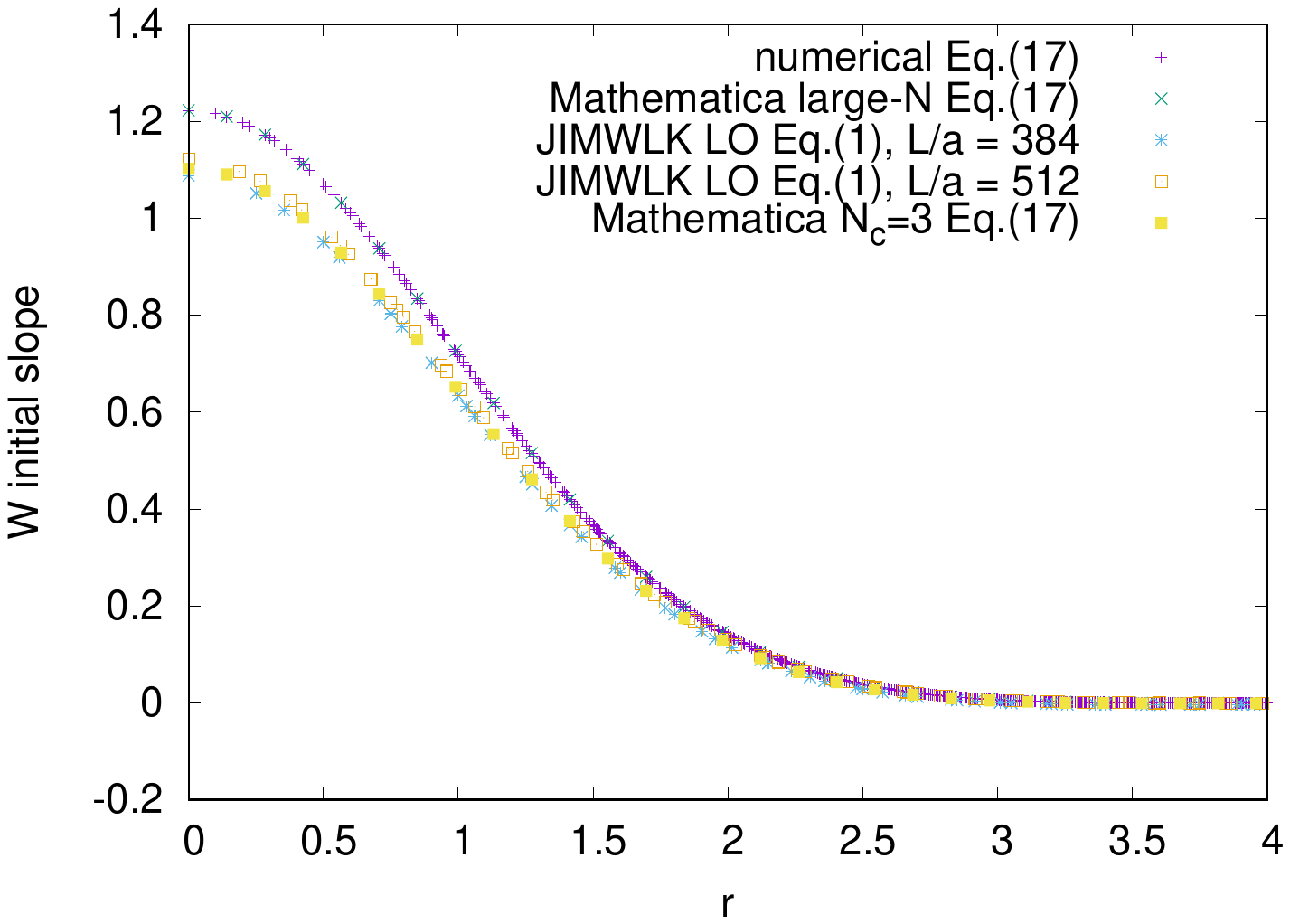}
    \caption{The initial slope of $W$ as a function of the distance $r=|\mathbf{x}-\mathbf{y}|$ obtained from the continuous integral performed by Mathematica, our discretized numerical estimate from Eq.\eqref{eq. W} and from the full simulation of Eq.\eqref{eq. JIMWLK}. \label{fig. 1}}
\end{figure}
\end{center}

It is instructive to investigate the initial behavior of the evolution equations Eqs.(\ref{eq. W}, \ref{eq. C}, \ref{eq. normal bk eq}). For simplicity, we choose a Gaussian initial condition, i.e.
\begin{multline}
    S(r,0) = \exp(-r^2/(2R^2_{\textrm{init}})), \\ \textrm{ and consequently } C(0) = 1, W(r,0) = \exp(-r^2/(2R^2_{\textrm{init}})).
\end{multline}
In the following we express all dimensional quantities in units of $R_{\textrm{init}}$ which is equivalent to setting $R_{\textrm{init}}=1$. In that case, we can evaluate the right-hand side of these evolution equations and calculate the initial slopes. The most insightful is the initial slope of $C$, $\frac{\partial C(\eta)}{\partial \eta}\Big|_{\eta=0}$.
\begin{equation}
    \frac{\partial C(\eta)}{\partial \eta}\Big|_{\eta=0} = \frac{\bar{\alpha}_s}{2 \pi} \int_{\mathbf{z}} \mathcal{K}_{xz} \Big(
        \exp(-|\mathbf{x}-\mathbf{z}|^2) -1 \Big).
\end{equation}
This has clearly a infrared divergence. In the numerical setup when the problem is discretized on a two-dimensional torus this divergence is regulated by the volume of the torus. Intuitively this may be expected, because as we make the torus larger and larger, the convolution gathers more and more noise which decorrelates $C$ faster. Note, that this divergence is a direct consequence of the infinite range of $\mathcal{K}_{xz}$ which emerges from the perturbative approximation for the gluon emission and does not hold in the complete theory with color confinement.

The expressions for the initial slope allow for an independent check of our numerics. As a test, we evaluate the continuous integral in Eq.\eqref{eq. W} using the Mathematica package and compare with our numerical discretized solution. We also include the results of the initial slope extracted from the full simulation of Eq.\eqref{eq. JIMWLK}. We present the results in Fig.\ref{fig. 1}. We observe that the semi-analytic results from Mathematica agree with our numerical solution. Both of them correspond to the large-$N_c$ limit and are about 10\% above the results from the full simulation of Eq.\eqref{eq. JIMWLK} as expected. We also include an additional check from Mathematica in which we have implemented into Eq.\eqref{eq. W} the non-local Gaussian approximation valid for finite $N_c$ which was derived in Ref.\cite{GAUSSIAN0_PhysRevLett.124.112301,GAUSSIAN2_Lappi_2016,GAUSSIAN3_Dumitru_2011,GAUSSIAN4_Dominguez_2009,GAUSSIAN5_Blaizot_2004} for equal-$\eta$ correlators of four Wilson lines. The detailed description of this correction is out of the scope of these proceedings, we just note that the correction eliminates the 10\% deviation between the full simulation and the large-$N_c$ result. We therefore conclude that we have our numerics under control.

\subsection{Full $\eta$ dependence}

\begin{center}
\begin{figure}
    \includegraphics[width=1.0\textwidth]{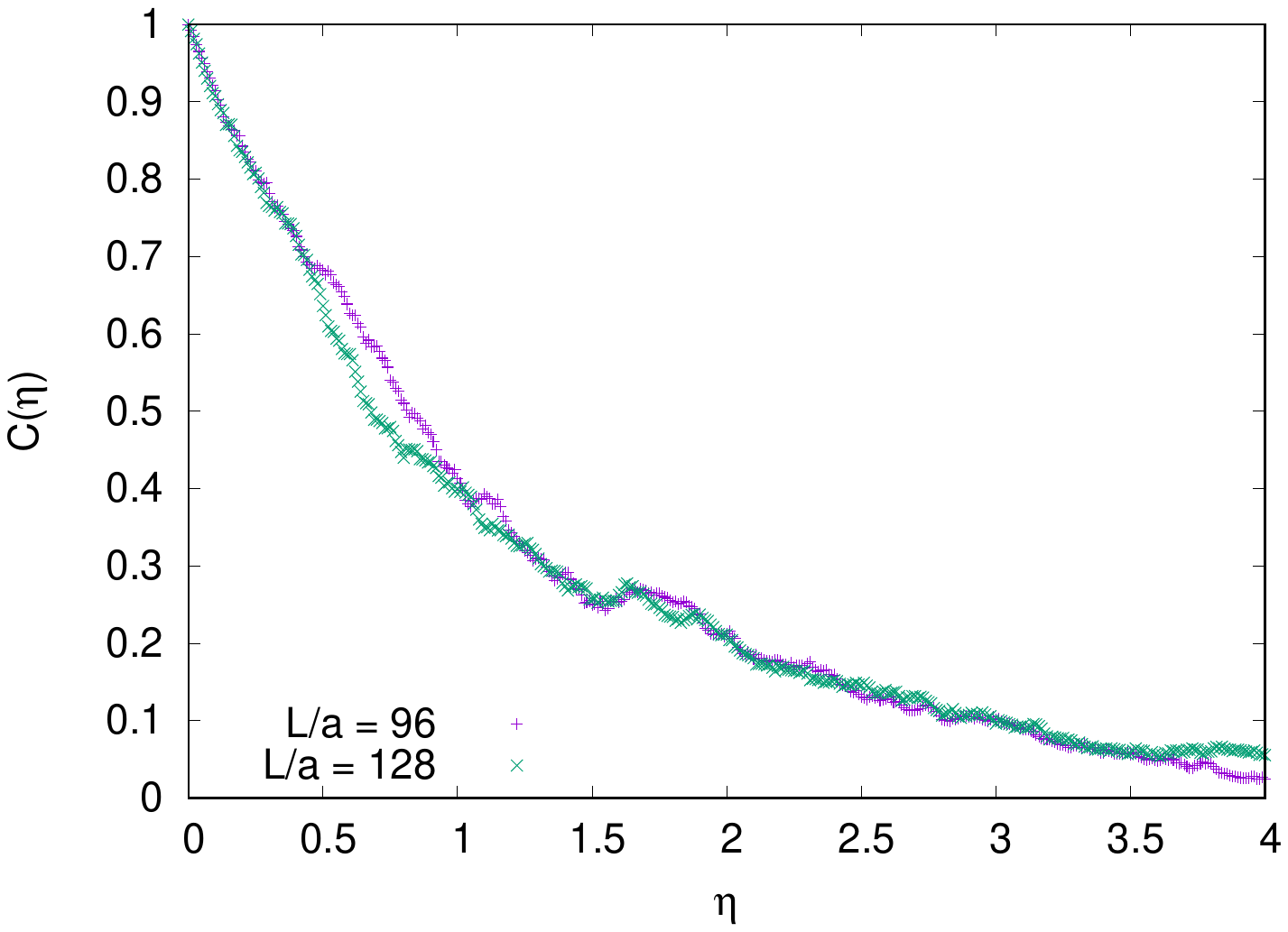}
    \caption{The correlation function $C(\eta)$ evaluated for two discretizations $L/a=96$ and $L/a=128$ with the volume of the torus set to $L=96 R_{\textrm{init}}$ where $R_{\textrm{init}}$ is the saturation radius of the initial condition. \label{fig. 2}}
\end{figure}
\end{center}

In Fig.\ref{fig. 2} we present the function $C(\eta)$ obtained from the full simulation of Eq.\eqref{eq. JIMWLK} at two different lattice spacings. We observe that both solutions yield results close to each other and that the function $C$ decays very quickly to zero. The simulations were performed in a volume of linear extent of $96 R_{\textrm{init}}$. Had we a larger linear extent, $C$ would decay faster. Consequently, the same is true for the $W(r, \eta)$ function which vanishes for large $\eta$. 

As it was stated in Section \ref{sec. recovering} for a certain range of arguments $\eta$, $r$ the full simulation of the JIMWLK equation with the kinematical constraint yields a dipole amplitude that evolves faster than the dipole amplitude obtained at leading order. This indicates that the dynamics of the improved JIMWLK equation does not reproduce the BK equation's expected dynamics with kinematical constraint. One of the necessary assumptions that allows to derive the improved BK equation from the improved JIMWLK equation is that the correlations function nonlocal in rapidity can be approximated by equal-rapidity correlation functions where both Wilson lines are taken at the smaller rapidity. For example, that would mean $C(\eta) \approx C(0) = 1$. This is not the case in full simulation as we have clearly demonstrated with Fig.\ref{fig. 2}. We note that the construction of the kinematically improved JIMLKW equation was derived in the continuum and on an infinite transverse plane. The above results might indicate that the translation of this construction onto the discretized torus with a finite volume which is typically used in the numerical setup may have some loophole that alters the dynamics in an undesired way.

\section{Conclusions}

In this contribution, we report on the progress in implementing the kinematical constraint into the Langevin formulation of the JIMWLK equation. The current implementation, however, leads to some unexpected results: the dipole amplitude extracted from the full numerical simulations of the improved JIMWLK equation does not reproduce the expected behavior predicted by the associated improved BK equation. The reduction of the improved JIMWLK equation to the BK equation relies on the fact that some correlation functions nonlocal in rapidity can be approximated by their local counterparts where both Wilson lines are taken at the same rapidity, the smallest of the two. We have proposed simplified examples of such correlation functions and derived their evolution equations. By solving numerically these evolution equations we found that the mentioned assumption is not fulfilled in our numerical setup, i.e. for example the correlation function $C(\eta)$ decays rapidly to zero whereas it was expected to be approximately equal to 1. Our results suggest that the role of correlation functions nonlocal in the rapidity variable has to be further investigated both in the BK equation and JIMWLK equation frameworks.

\section{Acknowledgements}
We gratefully acknowledge Polish high-performance computing infrastructure PLGrid (HPC Center: ACK Cyfronet AGH) for providing computer facilities and support within computational grants no. PLG/2022/015321 and no. PLG/2023/016656. T.S. kindly acknowledges the support of the Polish National Science Center (NCN) grant No. 2021/43/D/ST2/03375. P.K. acknowledges support from of the Polish National Science Center (NCN) grant No. 2022/46/E/ST2/00346.

\bibliographystyle{IEEEtran}
\bibliography{biblio.bib}

\end{document}